\documentstyle{article}
\begin{document} \title
{Constrained current algebras and $g/u(1)^d$ conformal field theories.}
\author{A.V.Bratchikov\\
Kuban State Technological University,\\
2,Moskovskaya St.,Krasnodar,350072,Russia}

\maketitle

\begin{abstract}
Operator quantization of the  WZNW theory invariant with
respect to an affine Kac-Moody algebra $\hat g$ with constrained $\hat
{u}(1)^d$ currents is performed using Dirac's
procedure.Upon quantization the initial energy-momentum tensor is
replaced by the $g/u(1)^d$ coset construction.The $\hat {su}(2)$ WZNW
theory with a constrained $\hat u(1)$ current is equivalent to
the $su(2)/u(1)$ conformal field theory.

\end{abstract}
{\bf 1.}In this article we study the connection between the WZNW theory
\cite{ew,kz}~, associated with a simply-laced algebra $g$,
and the $g/u(1)^d$ conformal field theory, whose energy-momentum
tensor $K_g$ is constructed via Goddard,Kent and Olive coset
method \cite{gko}
\begin{equation}
K_g=L_g-L_{u(1)^d}
\end{equation}
These theories for $d=r$,where $r=rank\> g,$ are closely related.  In
\cite{fz} Fateev and Zamolodchikov found that all the fields of the
$su(2)$ WZNW theory can be expressed in terms of the $su(2)/u(1)$
parafermions and a free boson.This relation and its generalization to
the case of arbitrary algebra g were used for computation of modular
invariant partition functions of $su(2)/u(1)$ \cite {gq} and
$g/u(1)^r$ \cite{dg} theories.

The object of this article is to present a new method of constructing
the $g/u(1)^d$,$1\le d \le r,$ conformal theory from the
WZNW theory invariant with respect to an affine Kac-Moody (KM) algebra
$\hat g$.  It is based on the operator quantization of the WZNW theory
with constraints \begin{equation}\label{constr} H_A(z)\approx 0,
\end{equation}
where
$H_A(z), A=1,\ldots ,d,$ are generators of $\hat{u}(1)^d$ subalgebra of the
algebra $\hat g$.Here and in what follows we treat only the holomorphic
part.
We quantize the system in terms of the modes of $H_A(z)$, which form an
algebra of first and second class constraints.The gauge is fixed
by an extention of this algebra.The theory can then be quantized
by replacing the initial operators by the operators which
are constructed using Dirac's procedure \cite{pd}.
This method can be generalized to other coset conformal theories.

{\bf 2.}The quantization will use the bosonic construction of the
KM currents and Virasoro generator $L_g$ of the WZNW theory in terms of
the Fubini-Veneziano fields
\begin{equation}\label{boson}
\varphi^j_s(z)=q^j_s-i{p^j_s}logz+i\sum_{n\not =0}{a_{sn}^j\over
n}z^{-n},
\end{equation}
where $s=1,\ldots,r,\> j=1,\ldots,k,\> k$ is the
level of the representation of the KM algebra $\hat g$ and
\begin{equation}\label{gamma}
[q_{s}^j,p_{t}^l]=i\delta^{jl}\delta_{st},\qquad
[a_{sm}^j,a_{tn}^l]=m\delta^{jl}\delta_{st}\delta_{m+n,0}.
\end{equation}
We denote by $\Gamma$ the algebra (4).
The bosons (\ref{boson}) have the two-point functions
\begin{equation}\label{two}
<0\vert \varphi_s^j(z)\varphi_t^l(w)\vert 0>=-\delta^{jl} \delta_{st}
\ln(z-w).
\end{equation}

The bosonic construction of the currents (\ref{constr}) at
level $k$ is given by
\begin{equation}
H_A(z)=\sum_{j=1}^k
i\partial_z\varphi_A^j(z).
\end{equation}
It is convenient to decompose the constraints into modes
\begin{eqnarray}
H_A(z)=H_{Ao}z^{-1}+\sum_{n\not= 0}{H_{An}z^{-n-1}},\nonumber\\
H_{Ao}=\sum^k_{j=1}{p^j_A},\qquad
H_{An}=\sum^k_{j=1}{a^j_{An}},
\end{eqnarray}
and consider an equivalent set
of constraints
\begin{equation}
H_{Ao}\approx0,\qquad H_{An}\approx0.
\end{equation}
The
operators $H_{Ao},H_{An}$ obey the KM algebra
\begin{equation} \label{class}
[H_{Ao},H_{Bo}]=[H_{Ao},H_{Bn}]=0,\qquad
[H_{Am},H_{Bn}]=km\delta_{AB}\delta_{m+n,0}.
\end{equation}

The commutator relations (\ref {class}) tell us that the operators
$H_{Ao}$ are first class constraints and the $H_{An}$ are second class
ones.  We take as gauge condition \begin{equation}\label{}
q_A\equiv\sum_{i=1}^k{q_{A}^i\approx 0}.
\end{equation}
The operators $q_{A}$ commute with $H_{Am}$ and
\begin{equation}\label{gauge}
[q_{A},H_{Bo}]=ik\delta_{AB}.
\end{equation}
It follows from (\ref{class}) and (\ref{gauge}) that the constraints
\begin{equation}\label{Phi}
\Phi=(H_{Ao},H_{An},q_A)
\end{equation}
are second class ones.

Let g be simply-laced.In this case the KM currents
$J(z)=(E_{\alpha}(z),H_s(z))$, where $\alpha$ are roots of
$g,s=1,\ldots,r,$ and energy-momentum tensor $L_g(z)$ of the WZNW
theory can be expressed in terms of $\Gamma$  using the
vertex operator representation \cite {fk,gs,dhs}.To find operators
which replace $J(z)$ and $L_g(z)$ in the constrained theory
it is sufficient to quantize only the operators $\Gamma$.

 Let $F$ and $G$ belong to the space $\Gamma$.According to Dirac's
procedure we replace these operators by the operators $\tilde F$ and
$\tilde G$ which satisfy the following commutator relation
\begin{equation}\label{brack} [\tilde F,\tilde
G]=[F,G]-\sum_{a,b}{[F,\Phi_a][\Phi_a,\Phi_b]^{-1}[\Phi_b,G]},
\end{equation}
where $\Phi=(\Phi_a)$ are the constraints (\ref{Phi}).
It is easy to see that all the commutators in the right-hand side
of (\ref{brack}) are c-functions. Therefore the new commutator is
well-defined.  Computations show that the nonvanishing commutators of
the operators
$\tilde{q}_{s}^j,\tilde{p}_{s}^j$ and $\tilde{a}_{sm}^j$
are given by
\begin{eqnarray}\label{}
[\tilde{q}_{A}^j,\tilde{p}_{A}^l]=i\eta^{jl},\qquad
[\tilde{q}_{I}^j,\tilde{p}_{I}^l]=i\delta^{jl},\nonumber
\end{eqnarray}
\begin{eqnarray}\label{commd}
[\tilde{a}_{Am}^i,\tilde{a}_{An}^j]=m\eta^{ij}\delta_{m+n,0},\qquad
[\tilde{a}_{Im}^i,\tilde{a}_{In}^j]=m\delta^{ij}\delta_{m+n,0},
\end{eqnarray}
where $I=d+1,\ldots,r$ and
\begin{equation}\label{}
\eta^{ij}=\cases{{(k-1)\over k}&if $i=j,$\cr
{-{1\over k}} &if $i\not=j.$ \cr}
\end{equation}
Note that $\eta^2=\eta$.It follows from (\ref{commd}) that the
constraints $\tilde{\Phi}=(\tilde{H}_{Ao},\tilde{H}_{An},\tilde{q}_A)$
commute with the operators $\tilde\Gamma=
(\tilde{q}_{s}^j,\tilde{p}_{s}^j,\tilde{a}_{sm}^j)$
\begin{equation}
[\tilde\Phi,\tilde\Gamma]=0
\end{equation}
The nonvanishing two-point functions of the fields
\begin{equation}\label{}
\tilde{\varphi}^j_s(z)=\tilde{q}^j_s-i\tilde{p}^j_slogz+i\sum_{n\not
=0}{\tilde{a}_{sn}^j\over n}z^{-n}, \end{equation}
are read
\begin{eqnarray} \label{twod}
 <0\vert
\tilde{\varphi}_A^i(z)\tilde{\varphi}_A^j(w)\vert
0>=-\eta^{ij}\ln(z-w),\nonumber\\
<0\vert
\tilde{\varphi}_I^i(z)\tilde{\varphi}_I^j(w)\vert
0>=-\delta^{ij}\ln(z-w),
\end{eqnarray}
where the vacuum vector is
defined by
\begin{equation}\label{}
\tilde{p}^i_s\vert 0>=0,\qquad \tilde{a}_{sn}^i\vert
0>=0\quad \hbox{for}\quad n>0
\end{equation}
It follows from (\ref{twod}) that the fields
$\tilde{\varphi}_s^j$ can be expressed as follows
\begin{equation}\label{subst}
\tilde{\varphi}^i_A=\eta^{ij} \varphi^j_A,\qquad
\tilde{\varphi}^i_I=\varphi^i_I
\end{equation}

Using the commutator relations (\ref{commd}) and the bosonic
construction of the currents and Virasoro generator of the WZNW
theory,one can compute operator product expansions of these operators
in the constrained theory.

These results can be generalized to non-simply-laced algebras using
the representation of the associated affine KM algebras in terms of the
operators $\Gamma$ and fermion fields\cite {gnos,bt}.

{\bf 3.}The energy-momentum tensor $L_g$ of the WZNW theory for g
simply-laced can by written as \cite{dhs}
\begin{eqnarray}\label{L}
L_g\left(\varphi
\right)={1\over{2(k+h)}}\{(1+h)\sum_{s=1}^r\sum_{j=1}^k:
(i\partial_z\varphi_s^j)^2:+\nonumber \\
+2\sum_{s=1}^r\sum_{i<j}^k:(i\partial_z\varphi_s^j)(i\partial_z\varphi_s^j):+
2\sum_\alpha\sum_{i<j}^k:\exp[i\alpha\cdot(\varphi^i-
\varphi^j)]:c^i_{\alpha}c^j_{-\alpha}\},
\end{eqnarray}
where $h$ is the dual Coxeter number of the algebra $g$,\ $c^i_{\alpha}$
are cocycles and the bosons $\varphi^i_s $ satisfy eq.(\ref{two}).

Let us consider this operator in the constrained theory where the fields
$\varphi^i_s $ are replaced by $\tilde{\varphi}^i_s $.Substituting
(\ref{subst}) into (\ref{L}), we get \begin{equation}\label{}
L_g(\tilde{\varphi})=
L_g(\varphi)-L_{u(1)^d}(\varphi),
\end{equation}
Thus the energy-momentum tensor of the constrained WZNW theory
can be written in the coset $g/u(1)^d$ form.

Consider the case of $su(2)/u(1).$\ $\hat{su}(2)$ algebra is generated
by the currents
\begin{equation}\label{curr}
E^+(z)=\sum_{j=1}^k{:e^{i\sqrt2\varphi^j(z)}:},\qquad
E^-(z)=\sum^k_{j=1}{:e^{-i\sqrt2\varphi^j(z)}:},
\end{equation}
where
$\varphi^i\equiv \varphi^i_1$ satisfy eq.(\ref{two}).  Upon
quantization of the system with the constrainted $\hat u(1)$ current
\begin{equation}\label{}
H(z)\equiv\sum^k_{j=1}{i\partial_z\varphi^j(z)}\approx 0,
\end{equation}
the fields $\varphi^i$ are replaced by
\begin{equation}\label{subst1}
\tilde{\varphi}^i=\eta^{ij}\varphi^j.
\end{equation}
Substituting (\ref{subst1}) into (\ref{curr}) we get
the parafermionic currents $\psi_1^+$ and  $\psi_1$
\begin{equation}\label{}
\sqrt k\psi_1^+ =
\sum_{j=1}^k{:e^{i\sqrt2\eta^{jl}{\varphi}^l}:},\qquad\sqrt
k\psi_1 = \sum_{j=1}^k{:e^{-i\sqrt2\eta^{jl}{\varphi}^l}:}.
\end{equation}
Computations show that $\psi_1^+$ and  $\psi_1$ satisfy the
parafermionic algebra of ref.\cite{fz}:
\begin{equation}\label{}
\psi_1(z)\psi_1^+(w)=(z-w)^{-2+{2\over
k}}\left(I+{{k+2}\over k}K_{su(2)}((w))(z-w)^2\right),
\end{equation}
where
$K_{su(2)}(z)=L_{su(2)}(\varphi(z))-L_{u(1)}(\varphi(z))$.
This proves the equivalence of the constrained $su(2)$ WZNW theory
to the $su(2)/u(1)$ theory.

In conclusion using canonical quantization techniques we have quantized
the WZNW theory invariant with respect to an affine KM algebra $\hat g$
with constrained $\hat u(1)^d$ currents.We have shown that Virasoro
algebra of the constrained theory is the $g/u(1)^d$ coset Virasoro
algebra.In the case of $su(2)/u(1)$ this correspondence also works for
the currents.It seems likely that the $g/u(1)^d$ conformal field
theory is equivalent to the constrained WZNW theory.It would be
interesting to study the $\hat g$ WZNW theory with
arbitrary constrained current algebra $\hat h \subset\hat g$ and compare
it with the $g/h$ conformal field theory.

\end{document}